\documentclass[%
 aip,
 amsmath,amssymb,
 reprint,%
groupedaddress]{revtex4-1}

\usepackage{graphicx}
\usepackage{dcolumn}
\usepackage{bm}
\usepackage[english]{babel}
\usepackage[utf8]{inputenc}
\usepackage[T1]{fontenc}
\usepackage{mathptmx}
\usepackage{etoolbox}

\makeatletter
\def\@email#1#2{%
 \endgroup
 \patchcmd{\titleblock@produce}
  {\frontmatter@RRAPformat}
  {\frontmatter@RRAPformat{\produce@RRAP{*#1\href{mailto:#2}{#2}}}\frontmatter@RRAPformat}
  {}{}
}%
\makeatother
\begin{document}

\preprint{AIP/123-QED}

\title{A Cryogenic Uniaxial Strain Cell for Quantum Devices}
\author{Bradley Lloyd}
\author{Davis Rash}
\author{Chandler Wilburn}
\author{Paul Kliewer}
\author{Meenakshi Singh}
\email{msingh@mines.edu}
\affiliation{%
 Department of Physics, Colorado School of Mines, 1500 Illinois St. Golden, Colorado, 80401 \\
}%

\date{\today}

\begin{abstract}
    Mechanical strain is a powerful resource for tuning quantum systems, but existing piezoelectric strain cells are generally optimized for fragile, high-aspect-ratio single crystals rather than the thick, square-profile chips typical of semiconductor quantum devices. Furthermore, adapting these cells for qubits requires accommodating dense RF and DC wiring while maintaining strict electrical isolation from high-voltage piezo actuators. Here, we present a piezoelectric uniaxial strain cell designed to homogeneously strain thick, square-profile substrates. We introduce a highly symmetric dual-chip loading configuration that effectively suppresses flexural deformation and shear stress. The cell integrates a high-density RF/DC interposer to support standard wire bonding and encloses the actuators in a grounded Faraday cage to prevent unwanted Stark shifts in the device layer. Finite element simulations confirm that combining stiff actuators with this symmetric mounting drastically improves strain homogeneity. Finally, we validate the apparatus experimentally by applying uniaxial strain to a 200 µm thick silicon die. Surface strain measurements demonstrate an applied strain of 215 $\mu\epsilon$ for 200 V applied piezo bias.
\end{abstract}

\maketitle


    Mechanical strain has emerged as a powerful control knob for tuning the electronic, magnetic, and structural properties of quantum materials.\cite{hicks_piezoelectric-based_2014,jo_uniaxial_2024} By breaking crystal symmetries and modifying bond angles and bond lengths, uniaxial strain can drive or enhance phase transitions, reshape Fermi surfaces, and selectively couple to nematic, superconducting, or density-wave order parameters. \cite{chu_divergent_2012,kim_charge_2021} Systematic use of uniaxial stress has, for example, revealed strong elastoresistive and elastocaloric signatures in correlated metals, \cite{ikeda_elastocaloric_2021} enabled controlled studies of nematic superconductivity, \cite{kostylev_uniaxial-strain_2020} and provided access to probe otherwise hidden phases in cuprates, ruthenates, and iron-based superconductors.\cite{kim_charge_2021,grinenko_unsplit_2021,ghosh_elastocaloric_2025}

    To exploit this tuning parameter at low temperatures, a family of cryogenic piezoelectric-based uniaxial strain and pressure cells has been developed over the past decade. In the design pioneered by Hicks and co-workers, piezoelectric stacks arranged in a compact, differential “three-stack” geometry provide continuously tunable compressive and tensile strain with high precision.\cite{hicks_piezoelectric-based_2014} The apparatus can be operated from room temperature down to dilution-refrigerator temperatures and is compatible with a wide range of probes.  Subsequent work has refined this approach to increase strain homogeneity and throughput, incorporate integrated force and displacement sensors, and to enable in situ operation in beamlines for angle-resolved photoemission spectroscopy,\cite{pfau_anisotropic_2021} x-ray scattering,\cite{kim_uniaxial_2018,kim_charge_2021,occhialini_spontaneous_2023} and neutron scattering.\cite{dashwood_strain_2023,sun_heisenberg_2021} Together, these developments have established piezoelectric-driven uniaxial strain as a standard tool in low-temperature condensed-matter experiments.


    In parallel, semiconductor spin qubits have progressed rapidly toward large-scale, fault-tolerant architectures, with silicon and Si/SiGe platforms achieving long coherence times, high-fidelity single- and two-qubit gates, and increasingly complex multi-qubit devices.\cite{neyens_probing_2024,george_12-spin-qubit_2025,wang_pursuing_2024} In these systems, strain is both an intrinsic design parameter and a promising resource: epitaxial strain in quantum wells and heterostructures sets valley splitting, modifies band offsets, and influences effective g-tensors, while controlled local strain can couple directly to donor or acceptor spin states, tune hyperfine interactions, and mediate spin–phonon coupling.\cite{zhang_acceptor-based_2023, oneill_engineering_2021, thorbeck_formation_2015} 
    
    Despite this opportunity, most existing uniaxial strain apparatuses are optimized for relatively fragile, high-aspect-ratio single crystals—typically millimeter-long bars with sub-millimeter cross-sections. These samples are mounted between anvils or glued onto slender mounting plates. Such geometries are ill-matched to the square or nearly square chips used in semiconductor quantum devices, where the lateral dimensions (a few millimeters on a side) are comparable to the thickness and the elastic modulus is high. 
    
   Beyond geometry, quantum-dot spin qubit experiments impose additional constraints: the strained chip must host dense RF and DC wiring, support microwave electric-dipole spin resonance (EDSR) drive lines, operate in a dilution refrigerator in large in-plane magnetic fields, and remain electrically isolated from the kilovolt-level piezo drive circuitry to avoid unintended Stark shifts and charge rearrangements. Past approaches to more advanced electrical characterization under the influence of strain have been limited to mounting devices directly on actuators,\cite{shayegan_low-temperature_2003,shayegan_two-dimensional_2006} destructive methods to reshape the devices into a geometry more amenable to strain, \cite{liu_continuously_2024} or probing bulk samples via microwave circuit resonance shifts. \cite{hosoi_strain-tunable_2025} 

    In order to bridge this gap between state-of-the-art quantum devices and strain cells, we present a strain cell capable of homogeneously straining quantum devices while still being fully compatible with the bonding and electrical characterization requirements that they impose.
    Such measurements present two challenges that, by the authors' knowledge, have not hitherto been solved in any strain cell design: (i) the propensity of square profile samples to flex and strain inhomogeneously and (ii) the difficulty in providing high density electrical contact (including RF contacts) to the sample under strain. 
        
\begin{figure}
        \includegraphics[width=3.4in]{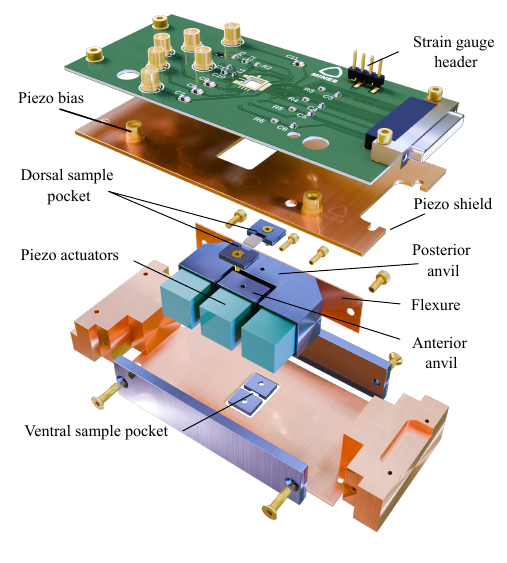}
        \caption{\label{fig:design} Exploded overview of the strain cell design.}
    \end{figure}

        The actuation scheme follows a three-stack, “Hicks-style” configuration: two outer piezoelectric stack actuators flank a central stack along the strain axis (Fig \ref{fig:design}). In operation the central stack is driven oppositely to the outer pair so that net displacement of the movable ``anvils" is differential with respect to the fixed frame, producing controllable tensile or compressive strain along the sample axis while maintaining a compact footprint compatible with superconducting magnets. 
    
        To efficiently strain wide, high-modulus device chips rather than narrow single-crystal beams, the design prioritizes actuator stiffness over maximum free stroke. The piezoelectric stacks therefore have a substantially enlarged cross-section compared to earlier cells, such that the stiffness of the actuator assembly exceeds that of the mounted chip. The stacks are bonded to molybdenum anvils whose large elastic modulus increases the overall stiffness of the moving bodies and thus improves strain transfer to the sample. Additionally, the coefficient of thermal expansion (CTE) of Mo ($5.4 \times 10^{-6} / \mathrm{K}$) is very well matched to the transverse CTE of the PZT actuators it is bonded to, thereby minimizing thermal strain applied to the actuators during cooldown.
    
        The anvils are guided relative to the cell body by a copper blade flexure. This flexure eliminates sliding contact and associated friction, and ensures reproducible motion during both actuation and thermal cycling. Because the three stacks are mounted such that their expansion along the poling direction during cooldown results in common motion, the net effect of their thermal expansion is a rigid translation of the anvils rather than unintended strain on the sample.

        The cell is designed to operate in a dilution refrigerator with an in-plane magnetic field up to $\pm 8~\mathrm{T}$. It is thus constructed from non-magnetic materials throughout. The main body is machined from oxygen-free copper and Mo, while all structural fasteners are brass. The larger thermal contraction of brass relative to copper and Mo provides an additional clamping preload on the internal components during cooldown. 

        A central feature of the design is the ability to apply homogeneous uniaxial strain to millimeter-scale, square or nearly square chips whose thickness is comparable to their lateral dimensions. In contrast to the high-aspect-ratio beams typically used in earlier uniaxial strain work, such chips are prone to flexural deformation and strain inhomogeneity under simple end loading. This problem is compounded by the need to mount the chip at the surface of the anvils such that high density wirebonding can be accomplished. This offset of the sample from the center of the actuators only increases the sample's propensity to bend.
    
        Conventional flexure designs proved ineffective in solving this, as they are not stiff enough to constrain the motion of a high-stiffness chip under the load forces required to strain it. We therefore introduce a flexure scheme in which the sample itself is half the flexure assembly: a second sample of equal stiffness is mounted on the ventral side of the anvils. This simultaneously allows the dorsal sample to be seated off-axis to be accessible to wire bonds while the bending moments that would otherwise cause a single chip to flex are largely canceled. Thus actuator motion is well constrained to the strain axis, shear stress on the actuators is greatly reduced, and the chips serve as their own flexures. 
        
        The sample mounts are realized as replaceable plates that are screw-attached to the strain anvils. Each plate is machined to form a recessed pocket that fully indents the chip. During mounting, the chip is placed in the pocket and an epoxy is applied so that it wets not only the bottom face but also creeps up the sidewalls of the chip. After cure, the chip is bonded along both its base and vertical side surfaces, substantially increasing the effective area near the device's top surface through which strain is transmitted.


        The strain cell is designed to support quantum-device-style wiring while the chip is under strain. A high-density RF/DC interposer is mounted directly above the strained chip, with bond pads located within roughly $1~\mathrm{mm}$ of the sample surface to allow standard gold wire or aluminum wedge bonding. The interposer provides $14$ DC lines and $5$ RF lines. Four of the RF lines are integrated with on-board bias tees, allowing combined DC and RF excitation up to $\sim 100~\mathrm{MHz}$ for gate and control signals. The remaining RF line is configured as an ESR drive line with a band-pass response with a $3.5$–$7.5~\mathrm{GHz}$ passband, suitable for microwave manipulation of spin qubits or other resonant excitations. A sixth coax connection sits on the board, to transmit the output of the cryogenic amplification circuit via un-attenuated steel coax line to a room temperature amplifier chain.
    
        A key requirement for quantum devices is the elimination of stray electric fields from the high-voltage piezo actuators at the device layer, to avoid unwanted Stark shifts and charge rearrangements. To this end, the actuator voltages are delivered from room temperature via un-attenuated stainless-steel coaxial cables terminated in shielded SMPM connectors that feed directly into the strain cell. These SMPM connectors are electrically isolated from the RF/DC interposer PCB. Within the cell, the piezo stacks are thus completely enclosed by a grounded copper shield. A copper cap plates over the top of the strain assembly, and an additional foil layer is placed beneath the sample mounts, so that the actuators are surrounded on all sides by conductive surfaces tied to a common ground. This geometry forms an electrostatic Faraday cage around the actuators and ensures that the quantum device sees only the intended gate and bias potentials.

    The strain cell's performance is first compared at room and base temperatures with no sample mounted to establish the loss in stroke the actuators (Piezomechanik PSt 150/10x10/7) suffer at dilution refrigerator temperatures. This is accomplished by monitoring the strain of one of the piezo actuators via an epoxied strain gauge. We find that the piezo actuators retain only 5.8\% of their room temperature stroke at 50 mK. This is a significant loss of stroke with temperature and can be overcome in future iterations by using materials like $SrTiO_{3}$ \cite{nasa_srtio3} or changing the composition of the PZT actuators to be less temperature dependent. \cite{pzt_temp} 
    
\begin{figure}
        \includegraphics[width=2.7 in]{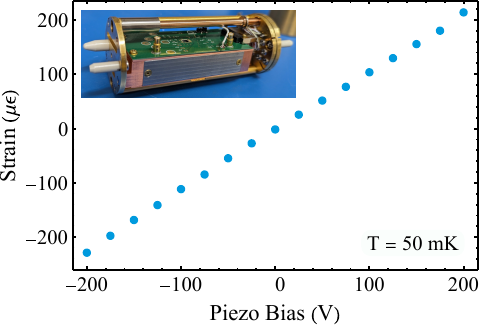}
        \caption{\label{fig:linear} Strain measured by the change in resistance of a 1.5 $\times$ 1.0 mm\textsuperscript{2} strain gauge epoxied to the center of the dorsal sample, showing a linear relationship between strain and voltage applied to the piezo stacks. Inset: assembled strain cell mounted to the dilution refrigerator puck.}
    \end{figure}

    The cell's ability to apply strain is validated by mounting two 2.7mm $\times$ 3.4mm $\times$ 0.2mm $\langle100\rangle$ silicon dies to the strain cell, one each to the dorsal and ventral sample plates. The dies are bonded via Stycast 2850 epoxy, with a measured bondline thickness of approximately 100 \(\mu\)m and are oriented so as to be strained in the [110] direction. The strain cell is mounted on the sample exchange puck of an Oxford Triton dilution refrigerator (inset, Fig. \ref{fig:linear}) and subsequently cooled to a temperature of 50 mK. The strain gauge, Micro-Measurements WK-13-031CF-350, is connected to a room temperature resistance bridge which is measured via standard lock-in techniques (Fig. \ref{fig:linear}). The response of the gauge is linear, yielding a slope of 1.085$\pm$0.005 \(\mu \)\(\epsilon \)/V. 
    
    We now compare the strain achievable with this setup against three relevant benchmarks: strains applied to specially prepared low-stiffness samples, strains achieved by directly epoxying samples to piezoelectric actuators, and theoretical strain requirements for observing phenomena of interest in spin-qubit devices. Previous studies using piezoelectric strain cells optimized for low-stiffness samples in long, narrow geometries have reported strains on the order of 10 millistrains.\cite{kostylev_piezoelectric-based_2019, barber_piezoelectric-based_2019} In the present work, an actuation voltage of up to $\pm 200$ V produces approximately $0.2$ millistrain on a conventional sample geometry. However, PZT actuators have been shown to withstand voltages as high as 600 V at cryogenic temperatures,\cite{kostylev_piezoelectric-based_2019} which would correspond to a maximum strain of approximately $0.6$ millistrain in the current setup without requiring specialized sample preparation. Compared with planar geometries in conventional cryogenic strain experiments, this stage delivers approximately $1.3$--$5\times$ greater strain per applied volt than the traditionally used technique of directly epoxying the sample to a piezo-actuator, used for similar geometries, which are typically implemented on substrates less stiff than silicon, such as GaAs.\cite{khisameeva_piezoplasmonics_2020,habib_anisotropic_2007} Furthermore, theoretical studies indicate that strains of only a few hundred microstrains are sufficient to begin exploring the predicted strain-tuned ``sweet spots'' in acceptor-based silicon quantum-dot qubit architectures.\cite{zhang_acceptor-based_2023}
    
    \begin{figure*}
        \includegraphics[width=5.5in]{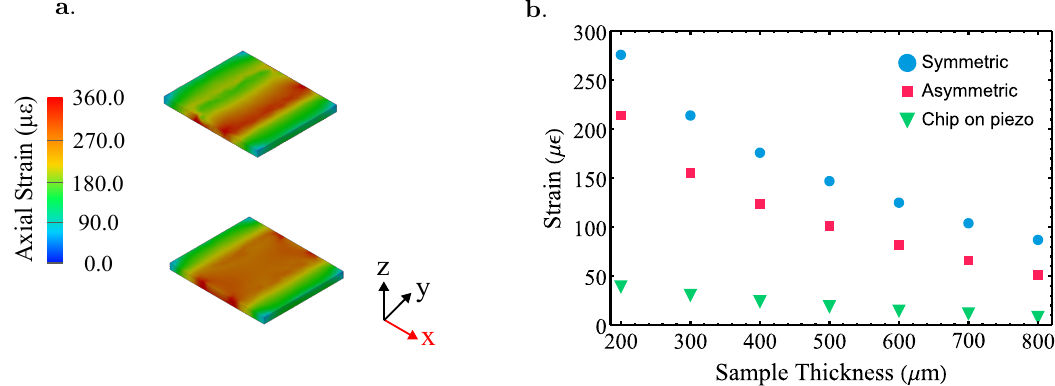}
        \caption{\label{fig:sims} Finite element analysis of axial strain (\(\epsilon \)\textsubscript{xx}) (a) Three-dimensional strain map of the dorsal silicon die under 200 V differential piezo bias at 50 mK, showing the distribution of axial strain across the chip surface for the asymmetrically loaded (top) and symmetrically loaded (bottom) cases. (b) Axial strain at the center of the 1×1 mm² region of interest as a function of substrate thickness for three mounting configurations: symmetric dual-chip loading (circles), asymmetric single-chip loading (squares), and chip epoxied directly to the actuator face (triangles). The symmetric configuration delivers superior strain homogeneity and maintains appreciable center strain even at substrate thicknesses up to 800 µm. }
    \end{figure*}

    Homogeneity of strain is a principal concern when straining samples with an aspect ratio close to one. While many devices patterned on the chip are so small that the gradient of applied strain is effectively zero in their vicinity, the measurement and calibration of the applied strain are only reliable if the macroscopic strain across the chip is sufficiently homogeneous. We therefore assess the strain distribution using three-dimensional finite element analysis (FEA) in NASTRAN, where the quasi-static operation of the piezoelectric actuators is modeled via a thermal-expansion analogy where the piezoelectric coefficients are mapped to the thermal expansion coefficients. \cite{cote_dynamic_2004,dong_dynamic_2006}

    The end of the strain anvils opposite the actuators is constrained to mimic the clamping condition of the cell frame, and the actuators are driven differentially at a bias of 200 V. All simulations utilize the measured 5.8\% actuation at 50 mK and a 100 \(\mu\)m thick bondline. The cryogenic material properties used are a Young's and Shear modulus of 16 GPa and 6 GPa respectively for the Stycast 2850,\cite{hicks_piezoelectric-based_2014,ojeda_temperature_2011} and Young's moduli of 400 GPa for Mo \cite{su_mechanical_2022} and 171 GPa for silicon strained along the [110] direction. \cite{liu_temperature-dependent_2021} All extracted strain values refer to the axial strain $\epsilon_{\text{xx}}$.

    The simulation cases considered are the asymmetric and symmetric loading conditions, in addition to a third technique commonly used for square profile devices where the sample is epoxied directly onto the face of an actuator.\cite{shayegan_low-temperature_2003,khisameeva_piezoplasmonics_2020,habib_anisotropic_2007} Because the sample is mounted off-center to facilitate electrical connection, the asymmetrically mounted chip experiences significant bending (Fig. \ref{fig:sims}a).    To mitigate this effect, we utilize the symmetric loading configuration in which a dummy chip with the same dimensions and elastic properties as the device chip is mounted in the ventral pocket. This balances the mechanical load on the actuators and forces the two strain anvils to move in a nearly parallel fashion under drive. Strain inhomogeneity is calculated as the difference between the maximum strain and the minimum strain within the 1$\times$1 mm\textsuperscript{2} center region  of interest relative to the strain in the center of this region:
    \begin{equation*}
        \Delta = \frac{\epsilon_{\text{xx}}^{\text{max}}-\epsilon_{\text{xx}}^{\text{min}}}{\epsilon_{\text{xx}}^{\text{center}}}\times100\%
    \end{equation*}
    The FEA results for this configuration show an improvement from 52.8\% deviation under the asymmetric loading condition to just 5.0\% deviation under the dual-chip, symmetric loading condition. Such homogeneity compares favorably to a simulated 8.7\% deviation for epoxying a chip directly onto the face of an actuator. The homogeneity is also necessary to ensure that the measured surface strain, which is an average over the area of the strain gauge, actually represents the strain on the quantum device. Since such devices only sit at the surface of their substrates, the inhomogeneity in the volume is irrelevant.
    
    Despite the fact that the stiffness of the load is doubled under symmetric loading, the strain at the center of the ROI is increased overall (Fig. \ref{fig:sims}b) as the strain is balanced across the region. The strain at the center of the sample is still appreciable even at sample thicknesses of 800 \(\mu\)m, a relatively large substrate thickness typical of some Si/Ge qubit platforms.\cite{george_12-spin-qubit_2025} Both cases provide significantly more strain than directly mounting a sample to the actuators as a result of the ``strain amplification" effect provided by actuating the anvils.

    In conclusion, we have simulated and measured a strain-cell architecture utilizing high-stiffness metallic components and actuators capable of generating hundreds of microstrains homogeneously on the surface of thick, square-profile substrates. Overall, the simulations predict \textasciitilde 20\% more  strain than what was measured, which we attribute to additional loss of strain in the epoxy, such as the epoxy used to bond the actuators to the anvils or the strain gauge to the sample. We have found that the cell can apply 1.085 \(\mu \)\(\epsilon \)/V to a 200 \(\mu \)m thick sample with an expected surface strain inhomogeneity of only 5\%, allowing the surface strain to be accurately measured via a conventional strain gauge.
    
    Crucially, this design maintains parity with standard wire-bonding and characterization protocols, ensuring seamless integration with the stringent environmental and measurement requirements of quantum devices. By bridging the gap between mechanical tuning and conventional quantum hardware architectures, this platform enables a new class of strain-dependent experiments, providing a critical pathway toward a deeper understanding of the operational environments of quantum devices.

\begin{acknowledgments}
This research was supported by the National Science Foundation under Award Number DMR-2046428. In addition, Bradley Lloyd was funded on an NRT fellowship under Award Number DGE-2125899. Some of the work was performed in Colorado School of Mines’ Shared Instrumentation Facility.
\end{acknowledgments}

\bibliography{references}

\end{document}